\documentclass[twocolumn,showpacs,fleqn,nobibnotes]{revtex4}

\usepackage{amsmath}
\usepackage{graphicx}
\usepackage{float}
\usepackage{subfigure}

\begin{document}

\title{Geometric scaling in inclusive charm production}
\author{V.P. Gon\c{c}alves
$^{1,a}$ \footnotetext{$^1$E-mail:barros@ufpel.tche.br}
and M.V.T. Machado $^{2,a,b}$\footnotetext{$^2$E-mail:magnus@if.ufrgs.br, magnus@ufpel.edu.br} }

\affiliation{$^a$ Instituto de F\'{\i}sica e Matem\'atica, Universidade Federal de
Pelotas\\
Caixa Postal 354, CEP 96010-090, Pelotas, RS, Brazil\\
$^b$ High Energy Physics Phenomenology Group, GFPAE,  IF-UFRGS \\
Caixa Postal 15051, CEP 91501-970, Porto Alegre, RS, Brazil}

\begin{abstract}
We show that the  cross section for inclusive charm production
exhibits  geometric scaling in a large range of photon
virtualities. In the HERA kinematic domain the saturation momentum
$Q_{\mathrm{sat}}^2(x)$ stays  below the hard scale
$\mu_c^2=4m_c^2$, implying charm production probing mostly the color
transparency regime and unitarization effects being almost  negligible. We derive our results considering 
two saturation models which  are able to describe the DESY $ep$ collider HERA data for the  proton structure function at small values of the Bjorken variable $x$.  A striking feature is the scaling on $\tau \equiv
Q^2/Q_{\mathrm{sat}}^2(x)$ above saturation limit, corroborating
recent theoretical studies.
\end{abstract}

\maketitle

{\it  Introduction.} The behavior of $ep/pp$ scattering in the
limit of high center-of-mass energy $\sqrt{s}$ and fixed momentum
transfer is one of the outstanding open questions in the theory of
the strong interactions. Over the past few years much theoretical
effort has been devoted towards the understanding of the growth of
the total scattering cross sections with energy. These studies are
mainly  motivated by the violation of the unitarity  (or
Froissart) bound by the solutions of the linear perturbative DGLAP \cite{DGLAP}
and BFKL \cite{BFKL} evolution equations. Since these evolution equations predict
that the cross section rises obeying a power law of the energy,
violating the Froissart bound \cite{FrMa}, new dynamical effects associated
with the unitarity corrections are expected to stop its further growth \cite{GLR,MUELLERS}. This expectation can be easily understood:
while for large momentum transfer $k_{\perp}$, the BFKL  equation
predicts that the mechanism $g \rightarrow gg$ populates the
transverse space with a large number of small size gluons per unit
of rapidity (the transverse size of a gluon with momentum
$k_{\perp}$ is proportional to $1/k_{\perp}$), for small
$k_{\perp}$ the produced gluons overlap and fusion processes, $gg
\rightarrow g$, are equally important. Considering the latter process,
the rise of the gluon distribution below a typical scale is reduced,
restoring the unitarity. That typical scale is energy dependent and is called  saturation scale  $Q_{\mathrm{sat}}$.  The saturation momentum sets the
critical transverse size for the unitarization of the cross
sections. In other words, unitarity is restored by including
non-linear corrections in the evolution equations \cite{GLR,MUELLERS,BARTELS,VENUGOPALAN,CGC,AYALAS,BAL,WEIGERT,KOVCHEGOV,LEVTUCH,BRAUN,LEVLUB,GBMS,TRIANTA}.
Such effects are small for $k_{\perp}^2 > Q_{\mathrm{sat}}^2$ and very strong for $k_{\perp}^2
< Q_{\mathrm{sat}}^2$, leading to the saturation of the scattering amplitude.
 The successful description of all
inclusive and diffractive deep inelastic data at the collider HERA
by saturation models \cite{GBW,BGBK,IPSAT}  suggests that these effects might become important
in the energy regime probed by current colliders.

One striking feature of the available saturation approaches  is
the prediction of the  geometric scaling. Namely,  the total
$\gamma^* p$ cross section at large energies is not a function of
the two independent variables $x$ and $Q$, but is rather a
function of the single variable $\tau = Q^2/Q_{\mathrm{sat}}^2$. As usual, $Q^2$ is the photon virtuality and $x$ the Bjorken variable. A current open
question is in what extent the geometric scaling is valid above
the saturation scale by studying the available high energy
formulations for the linear regime. In Ref. \cite{IANCUGEO} the
authors have demonstrated that the geometric scaling predicted at
low momenta $Q^2\leq Q_{\mathrm{sat}}^2(x)$ is preserved by the
BFKL evolution (at both fixed and running coupling constant)  up
to relatively large virtualities, within the kinematical window
$1\leq \ln \tau \ll \ln
(Q_{\mathrm{sat}}^2/\Lambda_{\mathrm{QCD}}^2)$. On the other hand,
in Ref. \cite{STASTOGEO}, the impact of the QCD DGLAP evolution on
the geometric scaling has been studied. In this case, the DGLAP
evolution equation is solved imposing as initial conditions along
the critical line $Q^2=Q_{\mathrm{sat}}^2(x)$ satisfying scaling,
showing that it is approximately preserved at very small $x$. The
residual scaling violation is factored out, although the determination of a window for the scaling above $Q_{\mathrm{sat}}$ is not
provided. As demonstrated in Ref. \cite{SGK}, the HERA data
on the proton structure function $F_2$ are consistent with scaling
at $x \leq 0.01$ and $Q^2 \leq 400$ GeV$^2$. Similar behavior has been
observed in exclusive \cite{MUNIERWALLON}  and  electron-nuclei processes \cite{SCALNUC}. These results,
while not entirely compelling, provide a strong motivation for
further investigations.

Here we  show that the data on inclusive charm production at HERA
exhibit geometric scaling above the saturation scale. Namely, the
charm inclusive cross section depends upon
$\tau=Q^2/Q_{\mathrm{sat}}^2(x)$ alone, where we have taken into
account  that the saturation scale is given by the saturation
model \cite{GBW}, i.e. $Q_{\mathrm{sat}}^2=(x_0/x)^{\lambda}$.  Moreover, we extend the symmetric saturation
model proposed in Ref. \cite{MUNIER} for the charm case and derive
an analytical expression for the charm production cross section,
which explicitly presents geometric scaling. We shown that both
prescriptions give similar results.

{\it Geometric scaling.} Lets consider the deep inelastic
scattering in the dipole frame, in which most of the energy is
carried by the hadron, while the virtual photon $\gamma^*$ has
just enough energy to dissociate into a quark-antiquark pair
before the scattering. In this representation the probing
projectile fluctuates into a
quark-antiquark pair (a dipole) with transverse separation
$r_{\perp}\sim 1/Q$ long after the interaction, which then
scatters off the proton. The interaction $\gamma^*p$ is further
factorized in the simple formulation \cite{PREDAZZI},
\begin{eqnarray}
\sigma_{L,T}^{\gamma^*p}(x,Q^2)=\!\int dz \,d^2r_{\perp}
|\Psi_{L,T}(z,r_{\perp},Q^2)|^2
\,\sigma_{dip}(x,r_{\perp}),\nonumber
\end{eqnarray}
where $z$ is the longitudinal momentum fraction of the quark,
$x\simeq Q^2/ W_{\gamma p}^2$ is equivalent to the Bjorken
variable.  The photon wave functions $\Psi_{L,T}$ are determined
from light cone perturbation theory and read as \cite{PREDAZZI}
\begin{eqnarray}
 |\Psi_{T}|^2 & = &\!  \frac{6\alpha_{\mathrm{em}}}{4\,\pi^2} \,
 \sum_f e_f^2 \, \left\{[z^2 + (1-z)^2]\, \varepsilon^2 \,K_1^2(\varepsilon \,r_{\perp})\right. \nonumber \\
& &  \left. +\,  m_f^2 \, \,K_0^2(\varepsilon\,r_{\perp})
 \right\}\label{wtrans}\\
 |\Psi_{L}|^2 & = &\! \frac{6\alpha_{\mathrm{em}}}{\pi^2} \,
\sum_f e_f^2 \, \left\{Q^2 \,z^2 (1-z)^2 \,K_0^2(\varepsilon\,r_{\perp})
\right\},
\label{wlongs} \nonumber
 \end{eqnarray}
where the auxiliary variable $\varepsilon^2=z(1-z)\,Q^2 + m^2_f$
depends on the quark mass, $m_f$. The $K_{0,1}$ are the McDonald
functions and the summation is performed over the quark flavors.

The dipole hadron cross section $\sigma_{dip}$  contains all
information about the target and the strong interaction physics.
In the Color Glass Condensate (CGC)  formalism \cite{CGC,WEIGERT}, $\sigma_{dip}$ can be
computed in the eikonal approximation, resulting
\begin{eqnarray}
\sigma_{dip} (x,r_{\perp})=2 \int d^2 b_{\perp} \,\left[
1-\mathrm{S}\,(x,r_{\perp},b_{\perp})\right]\,\,,
\end{eqnarray}
where $S$ is the $S$-matrix element which encodes all the
information about the hadronic scattering, and thus about the
non-linear and quantum effects in the hadron wave function. The
function $S$ can be obtained by solving an appropriate evolution
equation in the rapidity $y\equiv \ln (1/x)$. The main properties
of $S$ are: (a) for the interaction of a small dipole ($r_{\perp}
\ll 1/Q_{\mathrm{sat}}$), $S(r_{\perp}) \approx 1$, which characterizes that
this system is weakly interacting; (b) for a large dipole
($r_{\perp} \gg 1/Q_{\mathrm{sat}}$), the system is strongly absorbed which
implies $S(r_{\perp}) \ll 1$.  This property is associate to the
large density of saturated gluons in the hadron wave function. In
our analysis we will initially consider the  phenomenological
saturation model proposed in Ref. \cite{GBW} which encodes the
main properties of the saturation approaches. In this model
\begin{eqnarray}
\frac{\sigma_{dip}
(x,r_{\perp})}{\sigma_0}=1-\mathrm{S}\,(x,r_{\perp})\,;\,\,
\mathrm{S}=\exp\left[-\frac{Q_{\mathrm{sat}}^2(x)\,r_{\perp}^2}{4}\right],
\end{eqnarray}
with $\sigma_{dip}/\sigma_0$ the scattering amplitude, averaged
over all impact parameters $b_{\perp}$, and
$Q_{\mathrm{sat}}^2\simeq\Lambda^2\,e^{\lambda\ln(x_0/x)}$. The
parameters of the model were constrained from the HERA small $x$
data, coming out  typical values of order 1-2 GeV$^2$ for the
momentum scale. We have that when
$Q_{\mathrm{sat}}^2(x)\,r_{\perp}^2\ll 1$, the model reduces to
color transparency, whereas as  one approaches the region
$Q_{\mathrm{sat}}^2(x)\,r_{\perp}^2 \approx 1$, the exponential
takes care of resumming many gluon exchanges, in a
Glauber-inspired way. Intuitively, this is what happens when the
proton starts to look dark. It is important to emphasize that one moves  towards the unitarity bound for large $r_{\perp}$ in
this saturation model  much faster than the predicted by the CGC
approach. However, as our goal is the charm production, which is
dominated by the color transparency regime, the difference among
the approaches can be disregarded in what follows. The saturation
model depends upon the variables $x$ and $r_{\perp}$ only through
the dimensionless quantity $Q_{\mathrm{sat}}\,r_{\perp}$.
Consequently, the saturation model predicts the geometric scaling
of  the total cross section.

We are interested in the charm production in deep inelastic
scattering. From the  experimental point of view, the HERA
experiments have published data for the contribution of charmed
meson production to the structure function $F_2$. Their analysis
were based on $D^0$ and $D^*$ meson tagging. This allows one to
single out the charm contribution $F_2^c$ to the total structure
function and thus to investigate if the property of geometric
scaling is also present in this observable.  Before presenting our
results, lets perform a qualitative  analysis of the inclusive
charm production using the saturation model \cite{GBW} in order to shed light
on the dipole configurations dominating the process in the
relevant kinematical limits and show how the geometric scaling
comes out. A characteristic feature in heavy quark production
within the color dipole approach is that the process is dominated
by small size dipole configurations \cite{NIKZOLLER}. The overlap function
weighting the dipole cross section is peaked at $r_{\perp}\sim
1/m_c\simeq 0.1$ fm even for sufficiently low $Q^2$ values.  As a consequence, charm production is dominated
by color transparency and saturation effects are not important
there, i.e. $\sigma_{dip}\simeq \mathrm{S}_0\,(x,r_{\perp}) \equiv
\sigma_0\,Q_{\mathrm{sat}}^2(x) r^2_{\perp}/4$, which is the leading order term in the
expansion of $S$ for small $r_{\perp}$. We consider for simplicity
 the transverse contribution, whose corresponding wave function
is given by Eq. (\ref{wtrans}), and taking its asymptotic behavior
at  $\varepsilon\,r_{\perp}\ll 1$ such that
$K_1(\varepsilon\,r_{\perp})\sim 1/\varepsilon\,r_{\perp}$. This
imposes additional constraint to the $z$ integration, namely we
have the integrand should be multiplied by the Heaviside function
$\Theta(1-\varepsilon^2\,r_{\perp}^2)$, then
\begin{eqnarray}
\sigma_T^{c} & \sim  &  \int_0^{4/(Q^2+\mu_c^2 )}
\frac{dr_{\perp}^2}{r_{\perp}^2}
\mathrm{S}_0\,(x,r_{\perp})  \nonumber \\
& & \, + \,  \int_{4/(Q^2+\mu_c^2 )}^{4/\mu_c^2 }
\frac{dr_{\perp}^2}{r_{\perp}^2}\left(\frac{1}{Q^2r_{\perp}^2}
\right)\mathrm{S}_0\,(x,r_{\perp})\,, \label{semiqual}
\end{eqnarray}
where $\mu_c^2 = 4m_c^2$. The first term corresponds to the
symmetric dipole configurations, i. e. $<\! z \!> \approx 1/2$,  whereas
the second term comes from aligned jet configurations, with the
charm mass introducing a cut-off on the maximal size of the
charmed dipole.

For the HERA kinematical region, we have $Q_{\mathrm{sat}}^2 \approx 1 $
GeV$^2$, which implies that the relation $Q_{\mathrm{sat}}^2< Q^2 +
\mu_c^2$ is ever satisfied. Consequently, we can define two
kinematical regimes depending of the relation between $Q^2$ and
$\mu_c^2$. For $Q^2 \gg \mu_c^2$ we have scaling with logarithmic
enhancement coming from aligned jet configurations, whereas for
$Q^2 \ll \mu_c^2$ only symmetric dipole configurations contribute.
Therefore, we obtain that
\begin{eqnarray}
\sigma_T^{c} &  \sim &
\frac{\sigma_0\,Q_{\mathrm{sat}}^2(x)}{Q^2}
\left(1+ \ln \frac{Q^2}{\mu_c^2}  \right) \Theta (Q^2 - \mu_c^2) \nonumber \\
& &\, + \, \frac{\sigma_0\,Q_{\mathrm{sat}}^2(x)}{\mu_c^2}\Theta
(\mu_c^2-Q^2)\,,
\label{satexp}
\end{eqnarray}
where  the first term provides the behavior $1/\tau$ at large
$\tau$ whereas the second term leads to a smooth  transition down to
the  asymptotic ($\tau$-independent)  behavior at small $\tau$.

\begin{figure}
\includegraphics[scale=0.5]{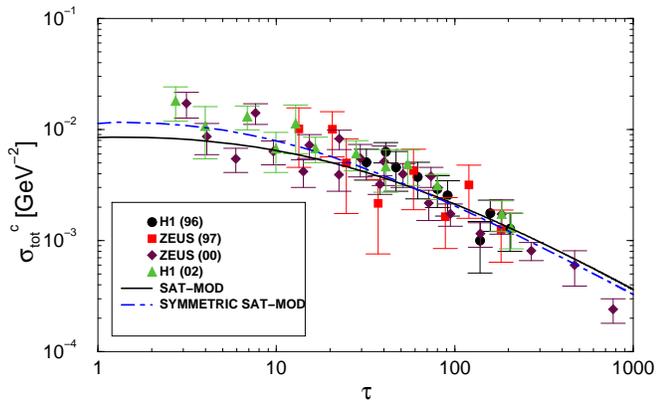}
\caption{Experimental data on inclusive charm
production plotted versus the scaling variable $\tau=Q^2/Q_{\mathrm{sat}}^2$.}\label{fig:1}
\end{figure}

An analytical expression for the $\tau$ dependence of the inclusive charm production  can also be obtained in a less model dependent way.  For this purpose we will make use of the symmetric saturation model \cite{MUNIER}, where the energy evolution of the proton leads to the parton multiplication and the transverse momentum scale $Q_{\mathrm{sat}}(x) $ appears. The main assumption is that the evolved proton can be described by a collection of independent dipoles at the time of the interaction whose sizes are distributed around $1/Q_{\mathrm{sat}}$. The rate of growth of the parton densities is assumed to be $Q_{\mathrm{sat}}^2(x)/\Lambda^2$ and the symmetry between low and high virtualities in $\gamma p$ interactions comes from the symmetry in the dipole-dipole cross section. The approach provides a quite intuitive and simplified expression for the inclusive production. Following such an  approach, we obtain  that in the HERA kinematical regime, i.e.  $Q_{\mathrm{sat}}^2<\mu_c^2$, the inclusive charm production is given by,
\begin{eqnarray}
\sigma^{c}_{tot}\,(\tau)= \frac{N_{c\bar{c}}}{\Lambda^2\,\nu_{>}}\,\left\{ 1- \exp \left[ -\frac{\nu_{>}}{\tau + \tau_c}\left( 1+ \log (\tau + \tau_c\right)     \right]\right\}\,, \nonumber
\end{eqnarray}
where $\tau_c= \mu_c^2/Q_{\mathrm{sat}}^2(x)$ and the parameters are taken from the data fit in Ref. \cite{MUNIER}. Here, we make the simplified assumption that the coupling with the dipole is flavor blind, in such way that $N_{c\bar{c}} = (2/5) N$, with $N$ being the global  normalization describing $F_2$ data. The factor $2/5$ corresponds  to the charge fraction $e_c^2/(\sum e_{u,\,d,\,s}^2 + e_c^2)$. Once the  parameters are fitted to proton structure function data, our prediction for the $\tau$ dependence in the inclusive charm production is parameter free. Moreover,   the expression above gives an analytical result for the whole  $\tau$ domain.

\begin{figure}
\includegraphics[scale=0.5]{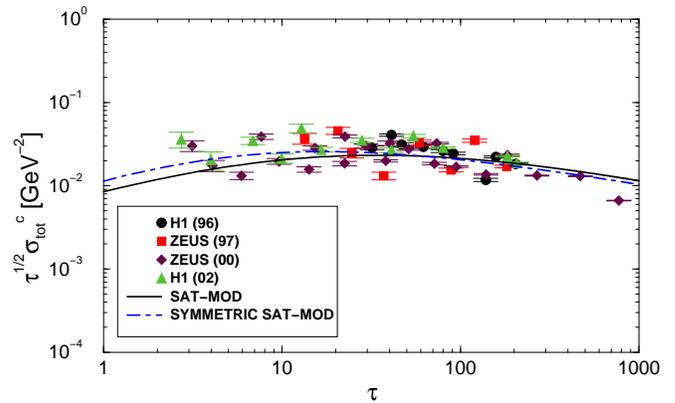}
\caption{The quantity $\sqrt{\tau}\,\sigma_{tot}$ plotted versus the scaling variable $\tau$.}
\label{fig:2}
\end{figure}

{\it Results.} In Fig.  \ref{fig:1} we show the experimental data
\cite{F2CDATA} on the total cross section for the inclusive charm
production plotted versus scaling variable
$\tau=Q^2/Q_{\mathrm{sat}}^2$, with $Q_{\mathrm{sat}}$  from the
saturation model. We include all available data covering the range
$1.5\leq Q^2\leq 130$ GeV$^2$ \cite{F2CDATA}.   It should be stressed that some bins with $x\geq 10^{-2}$ have also been included and old measurements contain  somewhat large uncertainties. Moreover, charm data are influenced by a  significant systematic uncertainty coming from the theoretical models used in the extraction of $F_2^{c\bar{c}}$. The combination of these factors produces a larger data dispersion than in the inclusive structure function.  We see the data exhibit geometric
scaling for the whole $Q^2$ range, verifying a transition in the
behavior on $\tau$  of the cross section from a smooth dependence
at small $\tau$ and an approximated $1/\tau$ behavior at large
$\tau$. The transition point is placed at $\mu_c^2=4m_c^2$, which
takes values of order 10 GeV$^2$ for a charm mass $m_c=1.5$ GeV.
This turns out in $\tau \simeq 10$ since at HERA
$Q_{\mathrm{sat}}^2\simeq 1$ GeV$^2$. The asymptotic $1/\tau$
dependence reflects the fact the charm production cross section
scales as $Q_{\mathrm{sat}}^2/Q^2$ modulo a logarithmic correction
$\sim \ln (Q^2/\mu_c^2)$, with energy dependence driven by the
saturation scale. The mild dependence at $\tau\leq \mu_c^2$
corresponds to the fact the cross section scales as
$Q_{\mathrm{sat}}^2/\mu_c^2$ towards  the photoproduction limit,
but with the same energy behavior given by the saturation scale.
As also plotted in Fig. \ref{fig:2}, we also found a symmetry
between the regions of large and small $\tau$ for the function
$\sqrt{\tau}\,\sigma^c_{tot}$ with respect the transformation $\tau
\leftrightarrow 1/\tau$ in the whole region of $\tau$. The features present in the inclusive charm production data can be well reproduced in the phenomenological saturation model as shown in Eq. (\ref{satexp}), corresponding to the solid curve in Figs.  \ref{fig:1} and  \ref{fig:2}. The symmetric saturation model also provides similar results, as shown in the dot-dashed lines. Disregarding the Glauber-like resummation in  this model, the  expression gets simplified to $\sigma^c_{tot}\propto \frac{1}{\tau+\tau_c}\,[1+\log(\tau + \tau_c)]$, and the  symmetric pattern is easily verified. Moreover, it is important to emphasize that a reasonable description of the charm photoproduction experimental data, which corresponds to $\tau = 0$,  is also  obtained using the symmetric saturation model.  Our results for the nuclear heavy quark photoproduction indicate that similar behavior is expected in the nuclear case \cite{VICMAGPHOTO}.

{\it Summary.}
Summarizing  our results, one  shows that the  inclusive charm
production  exhibits  geometric scaling in a large region  of
photon virtualities. In the HERA kinematic domain the saturation
momentum $Q_{\mathrm{sat}}^2(x)$ stays  below the hard scale
$\mu_c^2=4m_c^2$, implying that charm production probes mostly the color
transparency regime  and saturation corrections are not very important. In the
color dipole picture, the transition at $Q^2 \simeq \mu_c^2$ is
related to the presence of aligned jet configurations at $Q^2\gg
\mu_c^2$ and complete dominance of symmetric configurations at
$Q^2 \ll \mu_c^2$. An analytical result on the $\tau$ behavior was 
obtained within the symmetric saturation model, relying on quite 
simple assumptions about the dipole-proton interaction.  An outstanding  feature is the  scaling on $\tau$
above saturation scale, supported by recent theoretical
formulations.


\begin{acknowledgments}
The authors thank S. Munier for helpful comments and M.M. Machado for helping us in the preparation of the data plots. M.V.T.M. thanks the support of GFPAE IF-UFRGS, Porto Alegre. This work was  partially financed by the Brazilian funding
agencies CNPq and FAPERGS.
\end{acknowledgments}

\end{document}